\documentclass[preprint,aps,superscriptaddress,showpacs,nofootinbib]{revtex4-1}
\usepackage{geometry}

\usepackage{amscd}
\usepackage{amsmath,amssymb,amsthm,mathrsfs,amsfonts,dsfont}
\usepackage{subfigure, epsfig}
\usepackage{braket}
\usepackage{bm}
\usepackage{enumerate}

\usepackage{graphicx}
\usepackage{epsfig}
\usepackage{subfigure}
\usepackage{color}
\usepackage[10pt]{moresize}

\begin{document}

\title{1002 km Twin-Field Quantum Key Distribution with Finite-Key Analysis}

% Authors
\author{Yang Liu}
\affiliation{Hefei National Research Center for Physical Sciences at the Microscale and School of Physical Sciences, University of Science and Technology of China, Hefei 230026, China}
\affiliation{Jinan Institute of Quantum Technology and CAS Center for Excellence in Quantum Information and Quantum Physics, University of Science and Technology of China, Jinan 250101, China}
\affiliation{Hefei National Laboratory, University of Science and Technology of China, Hefei 230088, China}

\author{Wei-Jun Zhang}
\affiliation{State Key Laboratory of Functional Materials for Informatics, Shanghai Institute of Microsystem and Information Technology, Chinese Academy of Sciences, Shanghai 200050, China}

\author{Cong Jiang}
\affiliation{Jinan Institute of Quantum Technology and CAS Center for Excellence in Quantum Information and Quantum Physics, University of Science and Technology of China, Jinan 250101, China}
\affiliation{Hefei National Laboratory, University of Science and Technology of China, Hefei 230088, China}

\author{Jiu-Peng Chen}
\affiliation{Hefei National Research Center for Physical Sciences at the Microscale and School of Physical Sciences, University of Science and Technology of China, Hefei 230026, China}
\affiliation{Jinan Institute of Quantum Technology and CAS Center for Excellence in Quantum Information and Quantum Physics, University of Science and Technology of China, Jinan 250101, China}
\affiliation{Hefei National Laboratory, University of Science and Technology of China, Hefei 230088, China}

\author{Di Ma}
\affiliation{Jinan Institute of Quantum Technology and CAS Center for Excellence in Quantum Information and Quantum Physics, University of Science and Technology of China, Jinan 250101, China}

\author{Chi Zhang}
\affiliation{Hefei National Research Center for Physical Sciences at the Microscale and School of Physical Sciences, University of Science and Technology of China, Hefei 230026, China}
\affiliation{Jinan Institute of Quantum Technology and CAS Center for Excellence in Quantum Information and Quantum Physics, University of Science and Technology of China, Jinan 250101, China}

\author{Wen-Xin Pan}
\affiliation{Hefei National Research Center for Physical Sciences at the Microscale and School of Physical Sciences, University of Science and Technology of China, Hefei 230026, China}

\author{Hao Dong}
\affiliation{Hefei National Research Center for Physical Sciences at the Microscale and School of Physical Sciences, University of Science and Technology of China, Hefei 230026, China}
\affiliation{Jinan Institute of Quantum Technology and CAS Center for Excellence in Quantum Information and Quantum Physics, University of Science and Technology of China, Jinan 250101, China}

\author{Jia-Min Xiong}
\affiliation{State Key Laboratory of Functional Materials for Informatics, Shanghai Institute of Microsystem and Information Technology, Chinese Academy of Sciences, Shanghai 200050, China}

\author{Cheng-Jun Zhang}
\affiliation{Photon Technology (Zhejiang) Co. Ltd., Jiaxing 314100, China}

\author{Hao Li}
\affiliation{Hefei National Laboratory, University of Science and Technology of China, Hefei 230088, China}
\affiliation{State Key Laboratory of Functional Materials for Informatics, Shanghai Institute of Microsystem and Information Technology, Chinese Academy of Sciences, Shanghai 200050, China}

\author{Rui-Chun Wang}
\affiliation{State Key Laboratory of Optical Fibre and Cable Manufacture Technology, Yangtze Optical Fibre and Cable Joint Stock Limited Company, Wuhan, 430073, China}

\author{Chao-Yang Lu}
\affiliation{Hefei National Research Center for Physical Sciences at the Microscale and School of Physical Sciences, University of Science and Technology of China, Hefei 230026, China}
\affiliation{Hefei National Laboratory, University of Science and Technology of China, Hefei 230088, China}

\author{Jun Wu}
\affiliation{State Key Laboratory of Optical Fibre and Cable Manufacture Technology, Yangtze Optical Fibre and Cable Joint Stock Limited Company, Wuhan, 430073, China}

\author{Teng-Yun Chen}
\affiliation{Hefei National Research Center for Physical Sciences at the Microscale and School of Physical Sciences, University of Science and Technology of China, Hefei 230026, China}
\affiliation{Hefei National Laboratory, University of Science and Technology of China, Hefei 230088, China}

\author{Lixing You}
\affiliation{Hefei National Laboratory, University of Science and Technology of China, Hefei 230088, China}
\affiliation{State Key Laboratory of Functional Materials for Informatics, Shanghai Institute of Microsystem and Information Technology, Chinese Academy of Sciences, Shanghai 200050, China}

\author{Xiang-Bin Wang}
\affiliation{Jinan Institute of Quantum Technology and CAS Center for Excellence in Quantum Information and Quantum Physics, University of Science and Technology of China, Jinan 250101, China}
\affiliation{Hefei National Laboratory, University of Science and Technology of China, Hefei 230088, China}
\affiliation{State Key Laboratory of Low Dimensional Quantum Physics, Department of Physics, Tsinghua University, Beijing 100084, China}

\author{Qiang Zhang}
\affiliation{Hefei National Research Center for Physical Sciences at the Microscale and School of Physical Sciences, University of Science and Technology of China, Hefei 230026, China}
\affiliation{Jinan Institute of Quantum Technology and CAS Center for Excellence in Quantum Information and Quantum Physics, University of Science and Technology of China, Jinan 250101, China}
\affiliation{Hefei National Laboratory, University of Science and Technology of China, Hefei 230088, China}

\author{Jian-Wei Pan}
\affiliation{Hefei National Research Center for Physical Sciences at the Microscale and School of Physical Sciences, University of Science and Technology of China, Hefei 230026, China}
\affiliation{Hefei National Laboratory, University of Science and Technology of China, Hefei 230088, China}

\begin{abstract}
Quantum key distribution (QKD) holds the potential to establish secure keys over long distances. The distance of point-to-point QKD secure key distribution is primarily impeded by the transmission loss inherent to the channel. In the quest to realize a large-scale quantum network, increasing the QKD distance under current technology is of great research interest. Here we adopt the 3-intensity sending-or-not-sending twin-field QKD (TF-QKD) protocol with the actively-odd-parity-pairing method. The experiment demonstrates the feasibility of secure QKD over a 1002 km fibre channel considering the finite size effect. The secure key rate is $3.11\times10^{-12}$ per pulse at this distance. Furthermore, by optimizing parameters for shorter fiber distances, we conducted performance tests on key distribution for fiber lengths ranging from 202 km to 505 km. Notably, the secure key rate for the 202 km, the normal distance between major cities, reached 111.74 kbps.
\end{abstract}

\keywords{quantum key distribution, quantum communication, quantum optics, quantum information}
\maketitle

\section{Introduction}\label{sec_introduction}
Quantum key distribution (QKD) ~\cite{bennett1984quantum,ekert1991quantum,gisin2002quantum,scarani2009security,gisin2015far,xu2020secure,pirandola2020advances}  ensures secure key distribution using the principles of quantum mechanics. An active research frontier in practical quantum cryptography is exploring the distribution distance achievable with the present technology. The main challenge in extending the distribution distance of QKD is the channel loss that occurs when transmitting single-photon level quantum signals. Unlike classical communication, quantum signals cannot be amplified, posing a significant hurdle~\cite{wootters1982noncloning}. The exponentially decreased transmission of the optical fibre channel results in a vanishing secure key rate at long distances. Importantly, the limited number of detected signals over long distances also constrains the secure key rate due to the finite-key effect. 

In the quest for achieving long-distance QKD, twin-field QKD (TF-QKD) is proposed~\cite{lucamarini2018overcoming} recently. TF-QKD may achieve a secure key rate in spirit similar to that of a single-repeater QKD scenario, significantly increasing the ultimate distance. Within a few years, TF-QKD has already been experimentally demonstrated in lab~\cite{minder2019experimental,wang2019beating,liu2019exp,zhong2019proof,fang2020implementation,chen2020sending,liu2021field,chen2021twin,chen2022quantum,pittaluga2021600,wang2022twin,liu2023experimental} through up to 1002 km spooled fibre~\cite{liu2023experimental}, and in the field test over 511 km deployed fibre across metropolitans~\cite{chen2021twin}. However, the previous 1002 km distribution distance~\cite{liu2023experimental} was achieved under an asymptotic assumption. The longest distribution distance considering the finite size effect was reported to be 952 km~\cite{liu2023experimental}. The finite-size effect has to be taken into account in a practical QKD system since there are only a finite number of pulses. This consideration allows us to quantify the security level, i.e., the security coefficient; and apply the composable security framework. 

In this work, we demonstrate TF-QKD over 1002 km fibre channel considering the finite size effect. We adopt the sending-or-not-sending (SNS) protocol~\cite{wang2018sns} with advanced 3-intensity decoy-state method~\cite{hu2022universal} and the actively-odd-parity-pairing (AOPP)~\cite{xu2020sending, jiang2020zigzag} to improve the distribution distance. The ultra-low-noise superconducting nanowire single-photon detectors (SNSPDs) and dual-band phase estimation method are developed to suppress the system noise, thus achieving long distribution distance. Furthermore, the system is optimized for the normal distance between major cities of a fibre distance of 202 km. A secure key rate of 111.74 kbps is achieved at this distance, better than any reported results to our best knowledge.

\section{Results and Discussion}\label{sec_results}
\subsection{Protocol}\label{subsec_protocol}
In this work, we adopt the 3-intensity SNS protocol developed by Wang et al.~\cite{wang2018sns}. By placing the error correction process ahead of the decoy-state analysis process~\cite{hu2022universal}, we can utilize all the heralded time windows for decoy-state analysis, resulting in an enhanced key rate. Furthermore, we also apply the AOPP~\cite{xu2020sending} method to reduce the bit-flip error rate. The source parameters are symmetric for Alice and Bob: there are three sources on each side which are the vacuum source $v$, the decoy source $x$, and the signal source $y$ with intensities $\mu_v=0,\mu_x,\mu_y$ and probabilities $p_0,p_x,p_y$ respectively. In each time window, Alice (Bob) randomly prepares and sends out a pulse from the three candidate sources to Charlie who is assumed to measure the interference result of the incoming pulse pair and announce the measurement results to Alice and Bob. In this work, the raw keys in the time windows, while Alice and Bob choose the sources $v$ or $y$, are used to extract the secure keys. After Alice and Bob send $N$ pulse pairs to Charlie, and Charlie announces all measurement results, Alice and Bob distill the secure keys according to the following formula~\cite{jiang2020zigzag,jiang2021composable,hu2022universal}:
\begin{equation}
R=\frac{1}{N}\{n_1[1-H(e_1^{ph})]-fn_t H(E_t)\}-R_{\mbox{tail}},
\label{Eq:KeyRate}
\end{equation}
where $R$ is the key rate of per sending-out pulse pair; $n_1$ is the lower bound of the number of survived untagged bits after AOPP and  $e_1^{ph}$ is the upper bound of the phase-flip error rate of those survived untagged bits after AOPP; $n_t$ is the number of survived bits after AOPP and $E_t$ is the corresponding bit-flip error rate in those survived bits; $f$ is the error correction inefficiency which is set to $f=1.16$; $H(x)=-x\log_2x-(1-x)\log_2(1-x)$ is the Shannon entropy. And $R_{\mbox{tail}}$ is
\begin{equation}\label{r2}
\begin{split}
R_{\mbox{tail}}=\frac{1}{N}[2\log_2{\frac{2}{\varepsilon_{cor}}}+4\log_2{\frac{1}{\sqrt{2}\varepsilon_{PA}\hat{\varepsilon}}}\\+2\log_2(n_{vy}+n_{yv})],
\end{split}
\end{equation}
where $\varepsilon_{cor}$ is the failure probability of error correction, $\varepsilon_{PA}$ is the failure probability of privacy amplification, $\hat{\varepsilon}$ is the coefficient while using the chain rules of smooth min- and max- entropy~\cite{vitanov2013chain}, and $2\log_2(n_{vy}+n_{yv})$ is the extra cost of the advanced decoy state analysis~\cite{hu2022universal} ($n_{vy}$ is the number of raw keys while Alice chooses the source $v$ and Bob chooses the source $y$, and the definition of $n_{yv}$ is similar with that of $n_{vy}$ ).

\subsection{Experiment}\label{subsec_experiment}

\begin{figure*}[tbh]
\centering
\resizebox{14cm}{!}
{\includegraphics{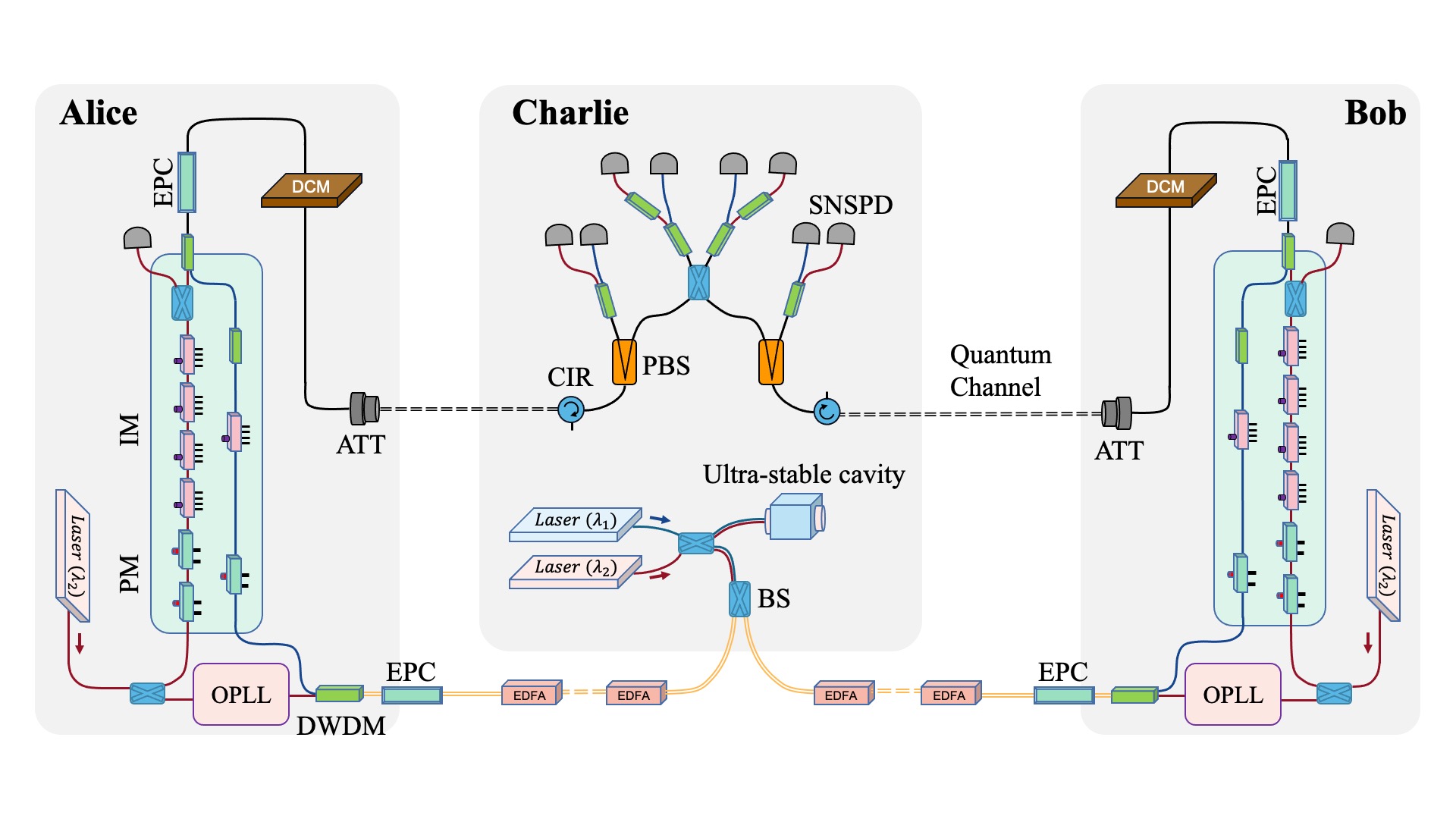}}
\caption{Experimental setup.
The seed lasers ($\lambda_1$=1548.51 nm and $\lambda_2$=1550.12 nm) for the phase reference and quantum signal are distributed from Charlie to Alice and Bob through 900 km single mode fibre spools.
The locally generated $\lambda_2$ light is frequency locked to the seed laser with an optical phase-locked loop (OPLL), and modulated to quantum signals. The ``dim phase reference'' is generated by time-multiplexing with the quantum signal. The $\lambda_1$ light is modulated to ``strong phase reference'', and then combined with the quantum signals by wavelength-multiplexing. 
The quantum signals are transmitted through the quantum channel and interfered at Charlie. The interference results are detected with SNSPDs. The local intensity monitors at Alice and Bob, the polarization and the relative delay at Charlie, and the ``strong phase reference'' signals are also detected with SNSPDs.
BS, beam splitter; PBS, polarization beam splitters; IM, intensity modulator, PM, phase modulator; ATT, attenuator; DWDM, dense wavelength division multiplexing; CIR, optical circulator; EDFA, erbium-doped fibre amplifier; DCM, dispersion compensation module; EPC, electronic polarization controller.
}
\label{Fig:TFQKD_Setup}
\end{figure*}

\begin{figure}[tbh]
\centering
\resizebox{12 cm}{!}
{\includegraphics{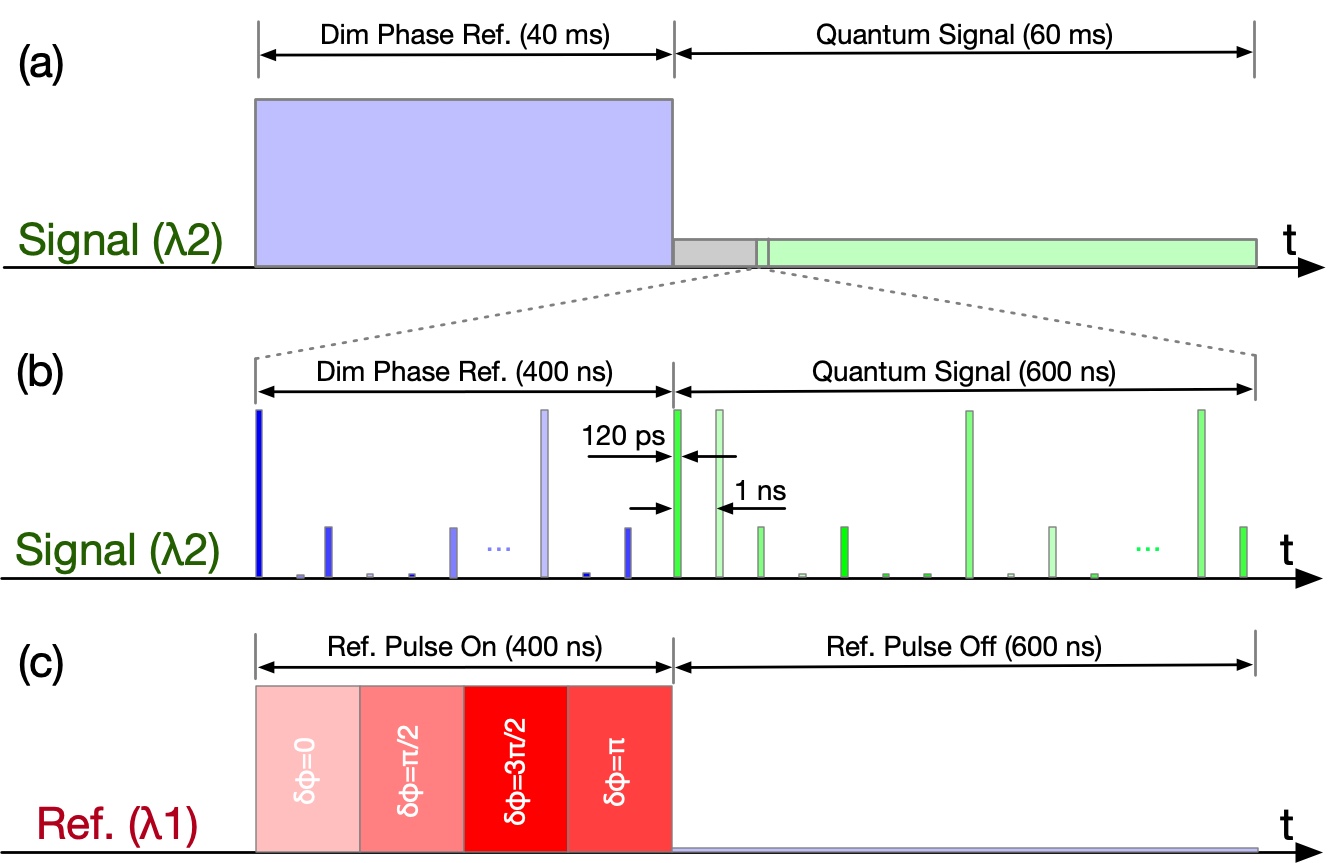}}
\caption{Time sequence of the quantum signals and the reference signals:
(a) The $\lambda_2$ light is modulated to a 100 ms period, including 40 ms ``dim phase reference'', and 60 ms ``quantum signals''. 
(b) The $\lambda_2$ light is modulated to a 1 GHz pulse train in each 1 $\mu$s period: in the first 400 ns, Alice and Bob modulate the pulses to the same intensity pattern, with four relative phases between them, as the ``dim phase reference''; in the remaining 600 ns, they modulate random ``quantum signals'' based on Alice's/Bob's random numbers, respectively. 
(c) The $\lambda_1$ ``strong phase reference'' light is modulated to a 400 ns pulse in each 1 $\mu$s period, with four relative phases between Alice and Bob. 
 }
\label{Fig:SignalSequence}
\end{figure}

The experimental setup is shown in Fig.~\ref{Fig:TFQKD_Setup}. The seed lasers ($\lambda_1$=1548.51 nm and $\lambda_2$=1550.12 nm) are frequency stabilized using the Pound-Drever-Hall (PDH) technique~\cite{pound1946electronic,drever1983laser, clivati2022coherent} with an ultra-stable cavity serving as the reference. The light is then sent to Alice's and Bob's stations through 450 km single-mode fibres respectively, incorporating 4 erbium-doped fibre amplifiers (EDFAs) in each path to stabilize the intensity. At Alice's (Bob's) station, the $\lambda_1$ light from Charlie is modulated to a 400 ns pulse in each 1 $\mu$s period, functioning as the ``strong phase reference''. The $\lambda_2$ light serves as the frequency reference for the optical phase-locked loop (OPLL). The locally produced $\lambda_2$ laser is locked to the frequency reference from Charlie, and then modulated to ``dim phase reference'' and ``quantum signals''. As shown in Fig.~\ref{Fig:SignalSequence}, the intensity of the 40 ms ``dim phase reference'' is generally higher than the 60 ms ``quantum signals'' light. In each 1 $\mu$s period, the pulse train is modulated to the same random pattern in the first 400 ns for Alice and Bob, serving as the ``dim phase reference''. In the remaining 600 ns, the pulse train is modulated to generate random quantum signals based on their respective local random numbers. Note that only the detections occurring within the first 400 ns in each 1 $\mu$s period, and within the first 40 ms in the 100 ms period, are utilized as the "dim phase reference". Similarly, the detections in the last 600 ns of each 1 $\mu$s period, and the last 60 ms of the 100 ms period, are used as the "quantum signals". For both the ``strong phase reference'' and ``dim phase reference'', the relative phases between Alice and Bob are set to four relative phases $\delta_{AB}=\{0, \pi/2, \pi, 3\pi/2\}$ within a 1 $\mu$s period.

We implemented the 3-intensity SNS-TF-QKD protocol in the experiment. The $\lambda_2$ quantum signals are modulated to 3 intensities with 16 different phases using intensity modulators (IMs) and phase modulators. The IMs are stabilized at Alice's (Bob's) station to ensure the quantum signals' intensities are stable. The $\lambda_1$ ``strong phase reference'' is then filtered and combined with the quantum signals. An electronic polarization controller is installed within the secure zone to control the polarization drift. A dispersion compensation module is employed to pre-compensate for the chromatic dispersion of the fibre channel. The signals are then attenuated to the predetermined intensities and subsequently transmitted to Charlie via the quantum channels.

The polarization of different wavelengths may evolute differently. At Charlie's measurement station, a polarization feedback algorithm is utilized to optimize the $\lambda_1$ detections to between 75 kHz and 300 kHz, while minimizing the $\lambda_2$. The relative delay between Alice's and Bob's signals is monitored and compensated using the rising edges of the $\lambda_1$ pulses. The light from Alice and Bob is interfered at the beam splitter and subsequently demultiplexed to $\lambda_1$ and $\lambda_2$ wavelengths. This light is then filtered by DWDMs, measured with SNSPDs, and recorded with a Time Tagger. The recorded signals are categorized into the ``strong phase reference'', ``dim phase reference'' and the quantum signal for subsequent data processing. 

The ultra-low loss fibre is utilized to minimize channel loss. The fibre is manufactured with ``pure silica core'' technology to reduce the doped Ge in the core and with decreased fictive temperature. The average attenuation of the fibres is measured to be less than 0.157 dB/km.

The ultra-low dark count SNSPDs are developed to reduce detection-related noise. The noise suppression includes stages of filtering. The long-wavelength ($>$2 $\mu$m) filtering is achieved using the 28 mm diameter fibre coils at the 40 K cold plate. Narrowband wavelength filtering is carried out utilizing a cryogenic bandpass filter (BPF) with a 5 nm bandwidth and an 85\% transmittance at 2.2 K cold plate~\cite{zhang2018fiber}. The dark count rate is measured to be as low as 0.02 Hz. Additionally, the detection efficiency is optimized to be around 60\% with a distributed Bragg reflector (DBR) based optical cavity~\cite{zhang2017nbn}. 

The time-multiplexed dual-band stabilization method is employed to reduce the re-Rayleigh scattering noise induced in previously reported time-multiplexed phase estimation procedures. With dual-band stabilization, the wavelength of the ``strong phase reference'' of $\lambda_1$ is different from the quantum signal of $\lambda_2$. The induced re-Rayleigh scattering is filtered with DWDMs. Furthermore, circulators are implemented in Charlie to eliminate noise resulting from the SNSPDs. Additionally, the $\lambda_1$ "strong phase reference" is time-multiplexed with the quantum signal, effectively mitigating disturbance caused by spontaneous Raman scattering noise induced by the strong phase reference light. The combination of wavelength- and time-multiplexing ensures that the noise introduced by the strong phase reference signal is less than 0.01 Hz. Moreover, the intensity of the weak "dim phase reference" signal, which is also time-multiplexed with the quantum signal, remains low enough to not generate perceptible noise.

The data post-processing-based phase estimation method \cite{liu2023experimental} is adopted. First, the phase drift of the $\lambda_1$ wavelength is estimated using the ``strong phase reference''. Then, the wavelength difference is taken into account to estimate the phase of the $\lambda_2$ light, with the accumulated phase drift of the ``strong phase reference''. Lastly, the initial phase difference $\phi_s(0)-\phi_r(0)$ is computed using the phase difference between the ``dim phase reference'' and the ``strong phase reference''. In the experiment, this phase difference is computed and refreshed every 500 ms, to circumvent any accumulation of errors stemming from inaccurate wavelength settings, high-order residual phase errors, and errors in phase estimation.

{\it Result.---}
\begin{figure}[tbh]
\centering
\resizebox{12 cm}{!}
{\includegraphics{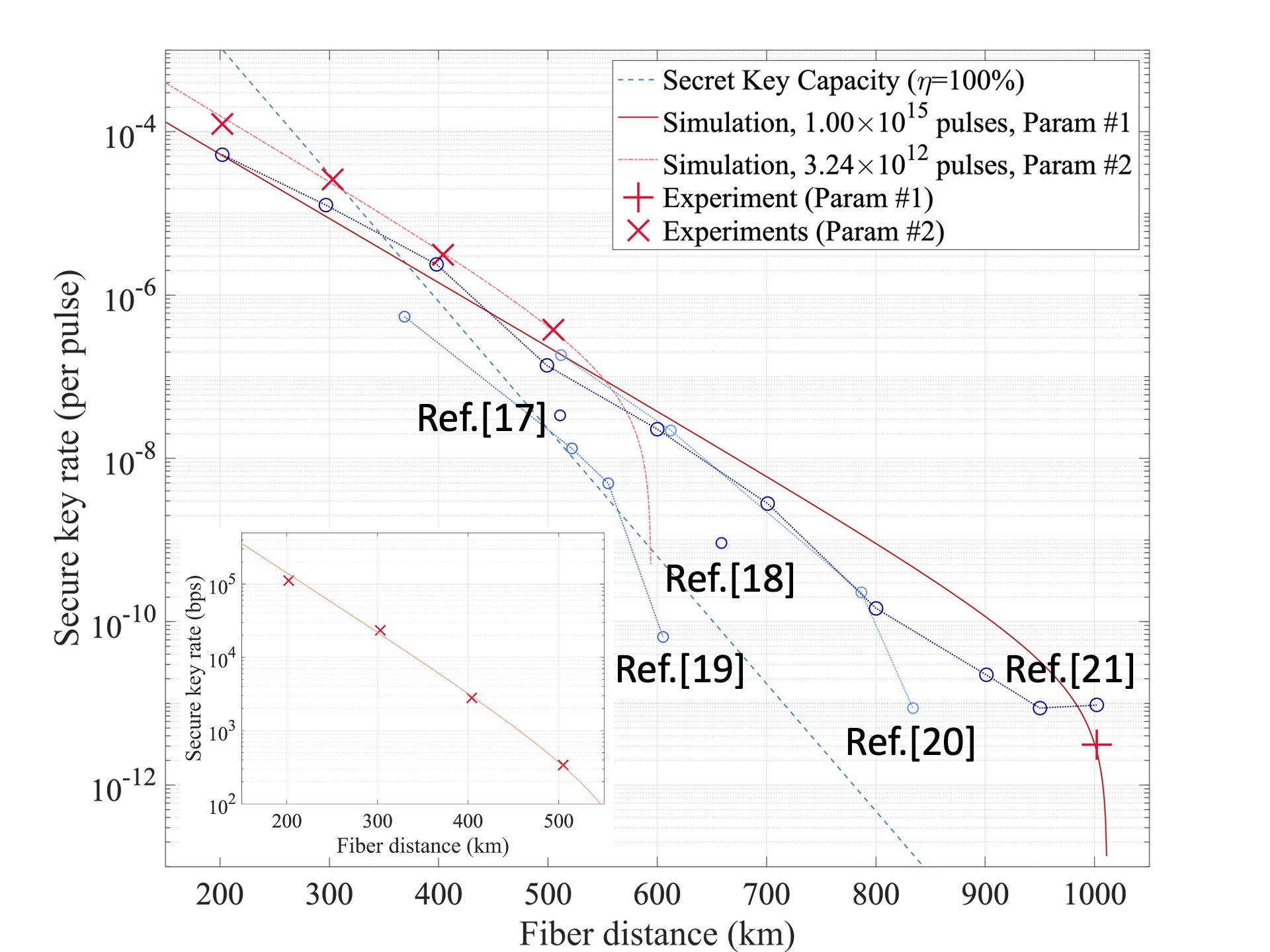}}
\caption{Simulations and experimental results of the secure key rates. The ``$+$''-shape points are experimental results using the long-distance optimized parameter (Parameter \#1), and the ``$\times$''-shape points are experimental results using the short-distance optimized parameter (Parameter \#2). The solid curve is the simulation results with Parameter \#1. The red dashed curve is the simulation result with Parameter \#2. All simulations and experimental results considered the finite size effect. The circle markers indicates the state-of-the-art TF-QKD results reported in Ref.~\cite{chen2021twin, pittaluga2021600, chen2022quantum, wang2022twin, liu2023experimental}. The blue dashed line shows the secret key capacity, i.e., the PLOB bound. Insert: the secure key rate per second in short distances.}
\label{Fig:KeyRate}
\end{figure}

\begin{table*}[htb]
\centering
  \caption{Experimental results for fibre lengths between 202 km and 1002 km. $L$: total fibre distance, $\eta$: total fibre transmittance, $N$: total signal pulses Alice and Bob sent, $n_1$: untagged bits after AOPP, $e_{1}^{ph}$: phase-flip error rate after AOPP, $n_t$: survived bits after AOPP, $E_t$: bit-flip error rate after AOPP, $E_x$: X-basis QBER of the raw bits, $R$: secure key rate per pulse, $R$ (bps): secure key rate per second.}
\begin{tabular}{c|ccccc}
\hline
$L$ 		& 202 km 	& 303 km 	& 404 km 	& 505 km	& 1002 km\\
$\eta$		& 31.6 dB	& 46.9 dB 	& 62.9 dB	& 78.6 dB	& 156.5 dB\\
\hline
$N$			& $3.24\times 10^{12}$ & $3.24\times 10^{12}$ & $3.24\times 10^{12}$ & $3.24\times 10^{12}$ & $1.00\times 10^{15}$\\
$n_1$		& $7.92\times 10^{8}$ & $1.56\times 10^{8}$  & $2.25\times 10^{7}$	& $3.29\times 10^{6}$	& 39454 \\
$e_{1}^{ph}$  	& 10.24\%  	& 9.17\% 	& 10.31\% 	& 13.24\%	& 17.05\%\\
$n_t$			& $2.17\times 10^{9}$   & $4.27\times 10^{8}$	& $6.22\times 10^{7}$  		& $9.41\times 10^{6}$ 		& 111671 \\
$E_t$ 	 		& $3.88\times 10^{-4}$ 	& $4.28\times 10^{-4}$ 	 & $2.28\times 10^{-3}$   	& $1.88\times 10^{-3}$   	& $9.44\times 10^{-3} $\\
$E_x$			& 4.25\%  	& 3.63\% 	& 3.76\%  	& 3.62\%	& 4.20\%  \\
\hline
$R$   		& $1.24\times 10^{-4}$ 	& $2.60\times 10^{-5}$ 	& $3.11\times 10^{-6}$    	& $3.76\times 10^{-7}$    	& $3.11\times 10^{-12} $\\
$R$ (bps) 		& 111,735 & 23,438 & 2,797 & 338 & 0.0011\\
\hline
\hline
\end{tabular}
\label{Tab:Result}
\end{table*}

We first test the performance of 1002 km fibres (the ``$+$''-shape points in Fig.~\ref{Fig:KeyRate}). The fibre distances between Alice-Charlie and Bob-Charlie are measured to be 500 km and 502 km. The decoy intensities are optimized as $\mu_x=0.08,\ \mu_y=0.445$, with the time ratios $p_{vac}=0.52,\ p_x=0.28,\ p_y=0.20$ (Parameter \#1). The finite size effect~\cite{jiang2021composable} is taken into consideration for all the experimental tests, considering composable security under any coherent attack~\cite{jiang2019unconditional, jiang2021composable}. The error correction inefficiency is set to $f=1.16$ in the calculation; the failure probability of Chernoff bound in finite-size estimation is set to $\varepsilon=10^{-10}$; the failure probability of the error correction, and the privacy amplification is set to $\varepsilon_{cor}=\varepsilon_{PA}=10^{-10}$; the coefficient of the chain rules of smooth min- and max- entropies is set to $\hat{\varepsilon}=10^{-10}$.

The system frequency is set to 1 GHz, with the signal pulse width set to 120 ps. The quantum signals are sent in the last 600 ns of the 1 $\mu$s period where the ``strong phase reference'' is switched off. The quantum signals are time-multiplexed with the ``dim phase reference'' in the last 60 ms of the 100 ms period. The detections near the strong light are also dropped to avoid potential noises. As a result, the effective signal frequency is 351 MHz for the long-distance scenario.

The total noises in the $\lambda_2$ are measured to be 0.019 Hz and 0.035 Hz in the working conditions. We attribute the noise mainly contributed by the SNSPD dark count and the spontaneous Raman scattering noise induced by the $\lambda_1$ light. The SNSPD detection efficiencies are measured to be 60\% and 55\%. The additional optical losses in Charlie are measured to be around 1.4 dB. In data processing, a 200 ps window is defined to filter out noises, with an efficiency of about 65\%.

Owing to the significant optical attenuation experienced over long-distance fibre, it is necessary to send a larger number of quantum signals in order to generate secure keys considering the finite size effect. A total of $1.00\times10^{15}$ quantum signal pulses are dispatched, resulting in $9.81\times10^{5}$ valid detections that fall within the effective window. The quantum bit error rate (QBER) in Z basis is measured to be $9.44\times10^{-3}$ after AOPP; the QBER in X basis is measured to be 4.20\%. The final secure key is $3.11\times10^{-12}$, which equates to 0.0011 bps considering the effective signal frequency. A total of 3112 bits of final secure keys are accumulated during the test. The detailed experimental results are summarized in Tab.~\ref{Tab:Result} and Fig.~\ref{Fig:KeyRate}. 

Next, we test the performance of fibre distances between 202 km and 505 km (the ``$\times$''-shape points in Fig.~\ref{Fig:KeyRate}). The intensities of the decoy states are optimized for short distances as $\mu_x=0.05,\ \mu_y=0.482$, with the time ratios $p_{vac}=0.68,\ p_x=0.04,\ p_y=0.28$ (Parameter \#2). The ``strong phase reference'' period is reduced to 100 ns in the 1 $\mu s$ signal period; the intensity of the ``dim phase reference'' is set to the same as the quantum signal through the 100 ms period. As a result, the effective signal frequency is increased to 900 MHz. Besides, we used SNSPDs with $>$80\% detection efficiency and a relatively higher dark count rate of about 10 Hz. The time window in data processing is set to 500 ps, yielding an almost unity efficiency.

A total of $3.24\times10^{12}$ quantum signal pulses are sent for each distance, which equals to one hour of experimental time. The secure key rate is measured to be $1.24\times 10^{-4}$, $2.60\times 10^{-5}$, $3.11\times 10^{-6}$, and $3.76\times 10^{-7}$ which corresponds to 111.74 kbps,  23.44 kbps, 2.80 kbps, and 338 bps for the 202 km, 303 km, 404 km, and 505 km fibre distances, respectively. The secure key rates exceed the absolute PLOB bound~\cite{pirandola2017fundamental} for the tests with the fibre distances equal to or longer than 404 km, where the PLOB bound is calculated as $-\log_2(1-\eta)$ with the optical and detection efficiency in Charlie set to $\eta_{opt}=100\%$.

\section{Conclusion}\label{sec_conclusion}

In conclusion, we have demonstrated the first experiment of SNS-TF-QKD over a remarkable distance of 1002 km, while considering the finite size effect. The result has been made possible by employing several key components, including the ultra-low-loss fiber, ultra-low-noise SNSPD, dual-band phase stabilization method, and moderate data size. The achieved secure key rates over fiber distances ranging from 202 km to 505 km were highly practical, indicating the potential for supporting a wide range of applications. In addition to improving the performance, TF-QKD is also expected to be implemented in chip-scale systems in the future, based on recent advancements in chip-scale systems implementing BB84 and MDI-QKD protocols~\cite{paraiso2021photonic,zhu2022experimental, wei2020high,cao2020chip}.

Y.L. and W.-J.Z. contributed equally.

\section*{Supplemental Material}
\subsection{Detailed Experimental Results}

The experimental results are summarized in Tab.~\ref{Tab:Result}.
In the table, we denote $N$ as the total number of signal pulses, $n_t$(After AOPP) as the remaining pairs after AOPP, $n_1$(Before AOPP) and $n_1$(After AOPP) as the number of the untagged bits before and after AOPP; $e_1^{ph}$(Before AOPP) and $e_1^{ph}$(After AOPP) as the phase-flip error rate before and after AOPP, $E_t$(Before AOPP) and $E_t$(After AOPP) as the bit-flip error rate before and after the bit error rejection by active odd parity pairing (AOPP). With all the parameters in the table, the final key rate per pulse and in one second is calculated as $R$ (per pulse) and $R$ (bps), $E_x$ as the phase-flip error rate of the sifted bits. We note that the ultra-low QBER E(After AOPP) allows us to use a practical error correction inefficiency $f=1.16$ in calculating the secure key rate.

In the following rows, we list the numbers of pulses Alice and Bob sent in different decoy states, labelled as ``Sent-AB'', where ``A'' (``B'') is ``0'', ``1'', or  ``2'', indicating the intensity Alice (Bob) has chosen within ``vacuum'', ``$\mu_x$'', or ``$\mu_y$''. With the same rule, the numbers of detections are listed as ``Detected-AB''. The total valid detections reported by Charlie is denoted as ``Detected-Valid-ch'', where ``ch'' can be ``Det1'' or ``Det2'' indicating the responsive detector of the recorded counts. The valid events falls in the preset Ds angle range is denoted as ``Detected-11-Ds'', the numbers of correct detections in this range is denoted as ``Correct-11-Ds''.

\linespread{1.2} 

\begin{table*}[htb]
\centering
  \caption{Experimental results for fiber lengths between 202 km and 1002 km (part I).}
\begin{tabular}{c|cccc|c}
\hline
$L$ 		& 202 km 	& 303 km 	& 404 km 	& 505 km	& 1002 km\\
$\eta$		& 31.6 dB	& 46.9 dB 	& 62.9 dB	& 78.6 dB	& 156.5 dB\\
\hline
$N$				& \multicolumn{4}{c|}{$3.24\times 10^{12}$}	& $1.00\times 10^{15}$\\
$R$ (per pulse) & $1.24\times 10^{-4}$ 	& $2.60\times 10^{-5}$ 	& $3.11\times 10^{-6}$	& $3.76\times 10^{-7}$	& $3.11\times 10^{-12} $\\
$R$ (bps) 		& 111,735 				& 23,438 				& 2,797					& 338 					& 0.0011\\
\hline
$n_1$(Before AOPP)	& $4.79\times 10^{9}$	& $9.69\times 10^{8}$ 	& $1.37\times 10^{8}$ 	& $2.02\times 10^{7}$	& 244481\\
$n_1$ (After AOPP)	& $7.92\times 10^{8}$	& $1.56\times 10^{8}$	& $2.25\times 10^{7}$	& $3.29\times 10^{6}$	& 39454\\
$n_t$				& $2.17\times 10^{9}$   & $4.27\times 10^{8}$	& $6.22\times 10^{7}$	& $9.41\times 10^{6}$	& 111671\\
$e_{1}^{ph}$(Before AOPP)	& 5.40\%	& 4.78\%	& 5.38\%	& 6.90\%	& 6.96\%\\
$e_{1}^{ph}$(After AOPP)  	& 10.24\%  	& 9.17\% 	& 10.31\% 	& 13.24\%	& 17.05\%\\
$E_t$ (Before AOPP)			& 29.05\%	& 29.12\%	& 29.17\%	& 29.17\%	&28.61\%\\
$E_t$ (After AOPP)	& $3.88\times 10^{-4}$ 	& $4.28\times 10^{-4}$	& $2.28\times 10^{-3}$	& $1.88\times 10^{-3}$	& $9.44\times 10^{-3} $\\
$E_x$				& 4.25\%  	& 3.63\% 	& 3.76\%  	& 3.62\%	& 4.20\%\\
\hline
\end{tabular}
\label{Tab:Result}
\end{table*}

\begin{table*}[htb]
\centering
  \caption{Experimental results for fiber lengths between 202 km and 1002 km (part II).}
\begin{tabular}{c|cccc|c}
\hline
Sent-00			& \multicolumn{4}{c|}{1497960000000}	& 271885442400000\\
Sent-01			& \multicolumn{4}{c|}{87480000000}		& 145295046000000\\
Sent-10			& \multicolumn{4}{c|}{88560000000} 		& 145629057600000\\
Sent-02			& \multicolumn{4}{c|}{617760000000} 	& 103877607600000\\
Sent-20			& \multicolumn{4}{c|}{616680000000} 	& 103543596000000\\
Sent-12			& \multicolumn{4}{c|}{35640000000} 		& 55946943000000\\
Sent-21			& \multicolumn{4}{c|}{36720000000} 		& 56280954600000\\
Sent-11			& \multicolumn{4}{c|}{5400000000} 		& 78993743400000\\
Sent-22			& \multicolumn{4}{c|}{253800000000} 	& 40582409400000\\\hline
Detected-Valid-Det1		& 5915818735	& 1196698346	& 170728132		& 25254691			& 591668\\
Detected-Valid-Det2		& 5897178099	& 1193249437	& 169543161		& 25212854			& 389346\\
\hline
% Note the data below are from 90deg
Detected-00			& 1878613		& 415261	& 317191	& 38651		& 2404\\
Detected-01			& 58792490		& 12385555	& 1721128	& 253911	& 53046\\
Detected-10			& 58225230		& 11297782	& 1681148	& 250398	& 52550\\
Detected-02			& 4003452009	& 841643692	& 115761488	& 17102689	& 199663\\
Detected-20			& 3928204193	& 761229174	& 112254967	& 16721359	& 198424\\
Detected-12			& 252858781		& 52869363	& 7324746	& 1079062	& 129903\\
Detected-21			& 257333673		& 50492830	& 7395166	& 1100667	& 130966\\
Detected-11			& 7310957		& 1464790	& 208135	& 31561		& 56925\\
Detected-22			& 3244940888	& 658149336	& 93607324	& 13889247	& 157133\\
\hline
Detected-11-Ds	 	& 1012651	& 203283	& 28908	& 5442	& 9858\\
Correct-11-Ds  		& 969577	& 195909	& 27820	& 5245	& 9444\\
\hline
\end{tabular}
\label{Tab:Result2}
\end{table*}

\bibliography{BibTFQKD}

\end{document}